\documentclass[english]{article}
\usepackage[T1]{fontenc}
\usepackage[latin9]{inputenc}
\usepackage{textcomp}
\usepackage{amsmath}
\usepackage{graphicx}
\usepackage{setspace}
\makeatletter
\usepackage[numbers, sort&compress]{natbib}

\makeatother

\usepackage{babel}
\begin{document}

\title{\noindent Multiparty Quantum Private Comparsion with Individually
Dishonest Third Parties for Strangers}

\author{Shih-Min Hung, Sheng-Liang Hwang, Tzonelih Hwang, and Shih-Hung Kao}
\maketitle
\begin{abstract}
This study explores a new security problem existing in various state-of-the-art
quantum private comparison (QPC) protocols, where a malicious third-party
(TP) announces fake comparison (or intermediate) results. In this
case, the participants could eventually be led to a wrong direction
and the QPC will become fraudulent. In order to resolve this problem,
a new level of trustworthiness for TP is defined and a new QPC protocol
is proposed, where a second TP is introduced to monitor the first
one. Once a TP announces a fake comparison (or intermediate) result,
participants can detect the fraud immediately. Besides, due to the
introduction of the second TP, the proposed protocol allows strangers
to compare their secrets privately, whereas the state-of-the-art QPCs
require the involved clients to know each other before running the
protocol. 

\end{abstract}

\section{Introduction}

Quantum private comparison (QPC) is an imperative branch of secure
multiparty computing, which allows participants to determine whether
their secrets are equal or not without revealing their secrets. The
first QPC protocol was proposed by Yang et al. \cite{1} using Einstein\textendash Podolsky\textendash Rosen
(EPR) pairs. The security in Yang et al.\textquoteright s protocol
is based on the use of decoy photons in the quantum transmission and
the one-way hash function for protection the secrets of the participants.
However, since the round trip quantum transmissions are adopted in
Yang et al.\textquoteright s protocol, special optical filters are
required to prevent Trojan horse attack \cite{2,3,4}, which decreases
the qubit efficiency. Accordingly, in order to enhance the qubit efficiency,
Chen et al. \cite{5} proposed a QPC protocol using a triplet Greenberger-Horne-Zeilinger
(GHZ) states. Since then, many QPC protocols \cite{6,7,8,9,10,11}
have been proposed based on various quantum entangled states. For
example, Tseng et al. \cite{11} proposed a QPC protocol without any
entangled EPR pairs and other QPC protocols such as in \cite{6,7,8,9,10}
use the EPR pairs, GHZ states, triplet W states and the $\chi$-type
genuine four particle entangled states for private information comparison.

The protocols described above can only compare the secrets for just
two participants. Until 2013, the first multiparty QPC protocol with
GHZ state was proposed by Chang et al. \cite{12}, in which $n$ participants
can compare whether the private information of any two users is equal
or not. Then Liu et al. \cite{13} proposed a multiparty QPC protocol
using d-dimensional basis state. Hereafter, many multiparty QPC protocols
have been proposed. Most of them also use the GHZ state or d-dimensional
basis state. Here, our proposed protocol is based on the GHZ state.

All the QPC protocols proposed so far require a third-party (TP) to
help the participants compare their secrets, generate photons and
announce the comparison (or intermediate) result. In this regard,
four types of QPCs can be categorized based on the levels of \textbf{trustworthiness
of the TP \cite{14}.}
\begin{enumerate}
\item First, TP is considered as an \textbf{honest} agent. Since the participants
can trust TP, they just send their secrets to TP for comparison. This
situation is an ideal one, but in reality, the assumption of an honest
TP is very unrealistic. 
\item Next, TP is considered as a \textbf{semi-honest} agent, where both
participants can trust TP partially. In this case, TP will loyally
execute the protocol, but may try to steal participants\textquoteright{}
secret using passive attacks. The semi-honest TP will passively collect
the classical information exchanged between participants and try to
reveal their secrets from this information.
\item Then, TP is considered as an \textbf{almost dishonest} agent, where
both participants can also trust TP partially. In this case, TP may
try to steal the information by modifying the procedure of the protocol
actively. However, it cannot collude with other participants. The
collude behavior includes the following cases: 
\begin{description}
\item [{(1)}] People works together to do something bad.
\item [{(2)}] A person helps the other person to avoid the detection if
he/she knows the other one is attacking.
\item [{(3)}] A person executes the protocol dependently with the other,
which should be independently in the protocol.
\end{description}
It means TP will not help any attacker steal the secrets of participants.
In other words, the TP not only can passively collect useful information
but also can actively perform any attack on the protocol except conspiring
with the participant. In some papers, this type of TP is also named
as semi-honest TP, a term easily confusing with the definition in
2.
\item Finally, TP is considered as a dishonest agent, where both the participants
cannot trust TP. This situation is the same as the two party QPC protocol
without TP, which has been proven to be insecure by Lo et al. \cite{15}.
\end{enumerate}

\subsection{Problem Statement and Motivation}

So far, we know that a TP plays a major role in many QPC protocols.
Even though several levels of trustworthiness of TP have been defined,
many recent QPC protocols adopt the assumption of an almost dishonest
TP, which unfortunately did not mention anything about whether or
not the TP will always announce a correct comparison (or intermediate)
result. However, if the TP announces a fake result, then all of above
protocols will be incorrect because the participants are not able
to detect this fraud. For example, if two participants are bidding
and comparing their prices, then the TP will announce a fake result
to disturb their bidding process, even if he/she cannot obtain useful
information and benefits. Hence, it is necessary for us to define
a new level of trustworthiness for this type of TP. Here, this particular
type of TP is called \textquotedblleft \textbf{individually dishonest
TP},\textquotedblright{} who could independently act maliciously.
The definition of individually dishonest TP is that the TP may announce
a fake result or try to actively steal the information by modifying
the procedure of the protocol except conspiring with participants
or other TPs. 

Hence, how to detect and prevent this individually dishonest TP\textquoteright s
malicious behavior is a challenging problem. The entire levels of
trustworthiness of TP can also be shown in Fig. 1, where except the
inner-most layer, each layer higher automatically assumes the capability
of the layer inner. 
\begin{figure}
\begin{centering}
\includegraphics[width=1\textwidth]{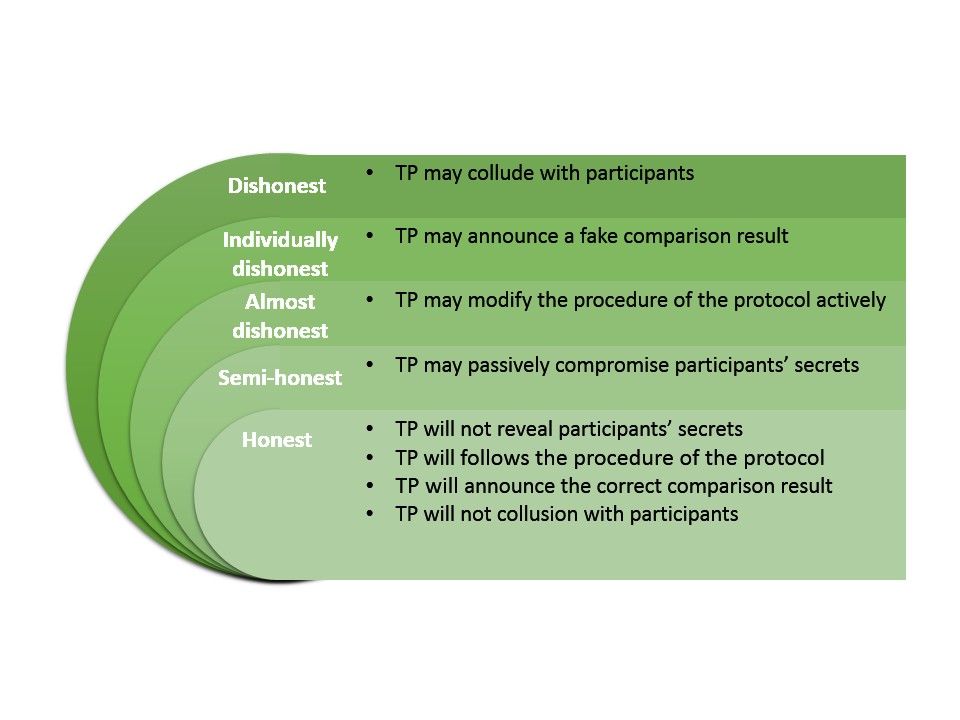}
\par\end{centering}
\caption{Level of trustworthiness of TP}
\end{figure}

According to the above figure, the individually dishonest is more
close to dishonest and hence is more practical, where except for the
conspiring attack, the TP can perform \textquotedblleft any\textquotedblright{}
possible attack \textendash{} including the denial-of-service attack
\textendash{} and the QPC protocol can still be secure. 

Furthermore, this article also investigates a new environment, called
the stranger environment, where participants could be strangers. As
contrary of this scenario, the state-of-the-art QPC protocols assume
the existence of authentication channels or pre-sharing keys between
participants in order to check the initial state or prevent private
information from leakage. The authentication channel allows the receiver
to conform the integrity of the transmitted message and the originality
of the sender, but the transmitted classical message is public. However,
sharing authentication channels or keys between the participants requires
them to establish some relationship beforehand. Can we construct a
QPC protocol for strangers who do not pre-share any key or quantum
states? To summarize our discussion, in this paper we intend to propose
a new multiparty QPC protocol with GHZ states, which is resilient
to the individually dishonest TPs in a stranger environment. 

In the following, we will consider Zhang et al.\textquoteright s protocol
\cite{14} as an example to show the problems with an individually
dishonest TP. Subsequently, a new QPC protocol will be proposed with
detailed security analysis.

The rest of this paper is organized as follows. Section 2 reviews
Zhang et al.\textquoteright s protocol and describes the problems.
Section 3 gives a solution protocol with individually dishonest TP
for strangers. Section 4 analyzes the security of the proposed protocol.
Finally, a concluding remark is given in Section 5.

\section{Zhang et al.\textquoteright s protocol and Problems}

Let Alice and Bob be two participants, who want to compare the equality
of their $m$-bit secret information $M_{A}$ and $M_{B}$ via the
help of an almost dishonest TP without leaking any private information
to the TP or any outsider. Zhang et al.\textquoteright s protocol
proceeds in the following steps:
\begin{description}
\item [{Step1}] TP prepares m EPR pairs randomly chosen from two Bell states
$\left|\phi^{+}\right\rangle $, $\left|\psi^{-}\right\rangle $,
where $\left|\phi^{+}\right\rangle =\frac{1}{\sqrt{2}}\left(\left|00\right\rangle +\left|11\right\rangle \right)$,
$\left|\psi^{-}\right\rangle =\frac{1}{\sqrt{2}}\left(\left|01\right\rangle -\left|10\right\rangle \right)$.
TP divides these EPR pairs into two sequences $S_{A}$ and $S_{B}$,
representing sequences of all the first photons and all the second
photons respectively.
\item [{Step2}] Step 2. TP prepares two sets of decoy photons $D_{A}$
and $D_{B}$ randomly chosen from $\left|0\right\rangle $, $\left|1\right\rangle $,
$\left|+\right\rangle =\frac{1}{\sqrt{2}}\left(\left|0\right\rangle +\left|1\right\rangle \right)$,
$\left|-\right\rangle =\frac{1}{\sqrt{2}}\left(\left|0\right\rangle -\left|1\right\rangle \right)$.
Each set contains m qubits. TP randomly inserts $D_{A}$ to $S_{A}$
(and $D_{B}$ to $S_{B}$) to form a new sequence $S_{A}^{*}$ (and
$S_{B}^{*}$), and then sends $S_{A}^{*}$and $S_{B}^{*}$ to Alice
and Bob, respectively.
\item [{Step3}] After Alice (Bob) receives $S_{A}^{*}$ ($S_{B}^{*}$),
she (he) and TP perform public discussion to check eavesdroppers. 
\item [{Step4}] After the public discussion, Alice and Bob can share many
Bell states and TP is the only one who knows the initial state of
these Bell states. Then, Alice, Bob and TP work together to check
the correctness of the states.
\item [{Step5}] Step 5. Alice (Bob) uses Z-basis to measure the photons
in $S_{A}$ ($S_{B}$). If the measurement result is $\left|0\right\rangle $,
then Alice (Bob) encodes it as the classical bit \textquoteleft 0\textquoteright ;
if the measurement result is $\left|1\right\rangle $, then Alice
(Bob) encodes it as the classical bit \textquoteleft 1.\textquoteright{}
Hence, Alice (Bob) obtains a key bit string $K_{A}$ ($K_{B}$).
\item [{Step6}] Alice (Bob) calculates the comparison information $C_{A}=K_{A}\oplus M_{A}$
($C_{B}=K_{B}\oplus M_{B}$), where $\oplus$ is a bitwise exclusive-OR
operation. They also collaborate together to compute the comparison
information $C=C_{A}\oplus C_{B}$ and send $C$ to TP.
\item [{Step7}] After TP gets $C$ from Alice and Bob, TP transforms the
initial Bell state $\left(S_{A},S_{B}\right)$ into a classical bit
string $C_{T}$ and calculates the comparison result $R=C_{T}\oplus C$.
If there is a \textquoteleft 1\textquoteright{} in $R$, then TP terminates
the protocol and announces the result that the two participants\textquoteright{}
secret information is different. Otherwise, (i.e., if all bits in
$R$ are \textquoteleft 0\textquoteright ), TP announces the result
that the two participants\textquoteright{} secret information is identical.
\end{description}
Within the protocol, if the TP announces a fake comparison result,
then according to Step 7, the participants cannot detect it. Hence,
the participants can do nothing but accept the wrong comparison result.
Besides, in Step 4, since the participants have to communicate with
each other to check the integrity of the almost dishonest TP so as
to avoid TP\textquoteright s manipulation of their communication,
they require to establish an authentication channel between them.
However, in a stranger environment, where both clients could be strangers
and hence do not share an authentication channel between them, this
protocol cannot be applicable. The same problems can also be found
in the other state-of-the-art QPC protocols such as in \cite{1,5,6,7,8,9,10,11,12,13,14}.

\section{The proposed scheme}

A multiparty QPC protocol for strangers with two individually dishonest
TPs is proposed here. Let $TP_{1}$, $TP_{2}$ be two individually
dishonest TPs. According to the previous definition, the individually
dishonest TPs may announce a fake comparison (or intermediate) result
to participants, though they cannot collude with each other or with
the participants. By the help of both TPs, participants can detect
whether any TP announces a wrong result. Besides, participants involved
in the protocol could be strangers, i.e., they do not need to pre-share
any secret or establish any authentication channel directly for communication
before-hand among them. In this protocol, there are quantum channels
and authentication channels between TPs and between each TP and each
participant. There are only classical channels between participants.

In this section, the GHZ states used in the protocol are first reviewed
in Section 3.1. The detail description of the proposed multiparty
QPC protocol is given in Section 3.2. The usefulness of the proposed
protocol in the stranger environment is described in Section 3.3.
Finally, the discussion about the malicious TP will be given in Section
3.4.

\subsection{The property of GHZ state}

The GHZ states are as follows:
\[
\left|\varPsi_{i}\right\rangle =\frac{1}{\sqrt{2}}\left(\left|q_{1},q_{2},...,q_{n}\right\rangle +\left(-1\right)^{\triangle}\left|\overline{q_{1},q_{2},...,q_{n}}\right\rangle \right),
\]
 where $i=1,2,3,...,2^{n}$, $q_{1}=0$, $q_{2},q_{3},...,q_{n}\in\left\{ 0,1\right\} $,
$\triangle=i-1$ (mod2) and $n$ denotes the number of participants.

The above state can also be re-written in X basis, $\left\{ \left|+\right\rangle ,\left|-\right\rangle \right\} $,
which is as follows:
\[
\left|\varPsi_{i}\right\rangle =\frac{1}{\sqrt{2^{n-1}}}\underset{n\left(-\right)=odd/even}{\sum}\left(-1\right)^{\delta}\left|x_{1},x_{2},...,x_{n}\right\rangle ,
\]
 where $x_{i}\in\left\{ +,-\right\} $ satisfies the condition of
$n\left(-\right)$, the number of $-$ in $x_{1},x_{2},...,x_{n}$.
If $\triangle=0$, then $n\left(-\right)$ will be even; otherwise,
if $\triangle=1$, then $n\left(-\right)$ will be odd. $\delta=\underset{\left\{ i|x_{i}=-\right\} }{\oplus}$.

For example, a three-qubit GHZ state $\left|\varPsi_{5}\right\rangle =\frac{1}{\sqrt{2}}\left(\left|010\right\rangle +\left|101\right\rangle \right)$
can be written in X-basis as follows:
\[
\begin{array}{lll}
\left|\varPsi_{5}\right\rangle  & = & \frac{1}{\sqrt{2^{3-1}}}\underset{even}{\sum}\left[\left(-1\right)^{\delta}\left|x_{1},x_{2},...,x_{n}\right\rangle \right]\\
 & = & \frac{1}{2}\left[\left(-1\right)^{0}\left|+++\right\rangle +\left(-1\right)^{1\oplus0}\left|+--\right\rangle +\left(-1\right)^{0\oplus0}\left|-+-\right\rangle +\left(-1\right)^{0\oplus1}\left|--+\right\rangle \right]\\
 & = & \frac{1}{2}\left(\left|+++\right\rangle -\left|+--\right\rangle +\left|-+-\right\rangle -\left|--+\right\rangle \right).
\end{array}
\]

According to Heisenberg uncertainty principle, the measurement result
of the $i-$th particle could be either $\left|q_{i}\right\rangle $
or $\left|\overline{q_{i}}\right\rangle $ with a probability of 50\%.
Hence, no one can predict the measurement result of the $i-$th particle.
However, for a particular GHZ state $\left|\varPsi_{w}\right\rangle $,
where $1\leq w\leq2^{n},$ if we measure two arbitrary particles,
e.g., the $i-$th particle and the $j-$th particle, and obtain the
measurement $m_{i}$ and $m_{j}$ respectively, then the xoring value
$m_{i}\oplus m_{j}$ is fixed. For example, let $n=4$ and $w=7$,
if the initial state is $\left|\varPsi_{7}\right\rangle =\frac{1}{\sqrt{2}}\left(\left|0011\right\rangle +\left|1100\right\rangle \right)$,
then the xoring value of first particle and second particle is always
\textquoteleft 0\textquoteright{} and the xoring value of second particle
and fourth particle is always \textquoteleft 1\textquoteright . Hence,
if one knows the initial state of a GHZ state, he can infer the xoring
value of measurement results of two arbitrarily particles. In the
following, TPs will utilize this property to do the comparison between
each pair of users. 

\subsection{Proposed Multiparty QPC protocol}

Let $P_{1}$, $P_{2}$, ..., $P_{n}$ denote $n$ participants, who
want to compare the equality of their m-bit secret information $M_{1}$,
$M_{2}$, ..., $M_{n}$ via the help of two individually dishonest
TPs, $TP_{1}$ and $TP_{2}$, without leaking any private information
to the TPs or any outsider. The proposed protocol proceeds in the
following steps: (as also described in Fig. 2)
\begin{description}
\item [{Step1}] $TP_{1}$ randomly prepares 2m n-particle GHZ states as
described in Section 3.1. $TP_{1}$ divides these GHZ states into
$n$ sequences $S_{i}$, where $1\leq i\leq n$, representing sequences
of all the $i-$th photons in these $2m$ initial states, respectively.
\item [{Step2}] $TP_{1}$ prepares $n$ sets of decoy photon $D_{1}$,
$D_{2}$, ..., $D_{n}$ randomly chosen from $\left|0\right\rangle $,
$\left|1\right\rangle $, $\left|+\right\rangle $, and $\left|-\right\rangle $.
Each set contains $2m$ qubits. $TP_{1}$ randomly inserts $D_{1}$
($D_{2}$, ..., $D_{n}$) into $S_{1}$ ($S_{2}$, ..., $S_{n}$)
respectively to form the new sequence $S_{1}^{*}$ ($S_{2}^{*}$,
..., $S_{n}^{*}$) and sends $S_{i}^{*}$ to $P_{i}$ respectively.
After $P_{i}$ receives $S_{i}^{*}$, he/she and perform the public
discussion to check the existence of eavesdroppers. First, $TP_{1}$
announces the positions and bases of decoy photons $D_{i}$. Then,
$P_{i}$ will divide $S_{i}^{*}$ into $S_{i}$ and $D_{i}$ by the
positions and use correct basis to measure the corresponding decoy
photon. Hereafter, the participants send back the measurement results
to $TP_{1}$. Finally, $TP_{1}$ checks the existence of eavesdroppers
by checking whether the measurement results are correct or not. If
they are correct, the protocol can be continued. Otherwise, the protocol
will be aborted. Then, $TP_{1}$ sends the information of the initial
GHZ states to $TP_{2}$ using quantum secure direct communication
protocol, e.g., \cite{16}.
\item [{Step3}] After the public discussion, $P_{1}$, $P_{2}$, ..., $P_{n}$
can share many GHZ states and $TP_{1}$ and $TP_{2}$ are the only
two who know the initial states of these GHZ states. Then, all participants
and $TP_{2}$ work together to check the correctness of the states.
For example, (1) $P_{1}$ randomly chooses the particles for checking
and announces the positions of those particles. (2) $P_{2}$ randomly
selects either Z-basis or X-basis for each chosen particle and announces
the bases. (3) All participants use the selected bases to measure
the corresponding particles and subsequently broadcast their measurement
results for each chosen particle. (4) $TP_{2}$ checks the measurement
results and the initial state sent from $TP_{1}$ and announces whether
or not the measurement results correspond with the initial states,
which should satisfy the equations described in Section 3.1. If yes,
then it implies that there is no eavesdropper and $TP_{1}$ prepares
the initial state loyally and also the information of initial state
sent from $TP_{1}$ is correct. Otherwise, they abort this protocol.
\item [{Step4}] $P_{i}$ uses Z-basis to measure the photons in $S_{i}$
and obtains a key string of measurement result $K_{i}$. That is,
if the measurement result is $\left|0\right\rangle $, then $P_{1}$
encodes it as the classical bit \textquoteleft 0\textquoteright .
If the measurement result is $\left|1\right\rangle $, then $P_{1}$
encodes it as the classical bit \textquoteleft 1\textquoteright .
calculates the comparison information $C_{i}=K_{i}\oplus M_{i}$.
\item [{Step5}] $P_{i}$ sends $C_{i}$ to $TP_{1}$ and $TP_{2}$ via
authenticated channels.
\item [{Step6}] After $TP_{1}$ gets $C_{i}$'s from all participants,
for arbitrary two participants, $P_{i}$ and $P_{j}$, $TP_{1}$ calculates
the comparison result $R_{ij}=T_{ij}\oplus C_{i}\oplus C_{j}$, respectively,
where $T_{ij}$ is the expected xoring value of the $i-$th and $j-$th
particles in that particular GHZ state. If there is a \textquoteleft 1\textquoteright{}
in $R_{ij}$, then $TP_{1}$ announces that the secret information
of $P_{i}$ and $P_{j}$ is different. Otherwise, $TP_{1}$ announces
that the secret information of f $P_{i}$ and $P_{j}$ is identical.
Similarly, $TP_{2}$ also does the comparison and announces the comparison
result, too.
\item [{Step7}] Any two participants, $P_{i}$ and $P_{j}$, can compare
the $R_{ij}$between $TP_{1}$ and $TP_{2}$. If the results are the
same, then they believe both $TP_{1}$ and $TP_{2}$ announce the
correct result. Otherwise, they know that one of TPs announce a wrong
result and the entire comparison process will be aborted.
\end{description}
\begin{figure}
\begin{centering}
\includegraphics[width=1\textwidth]{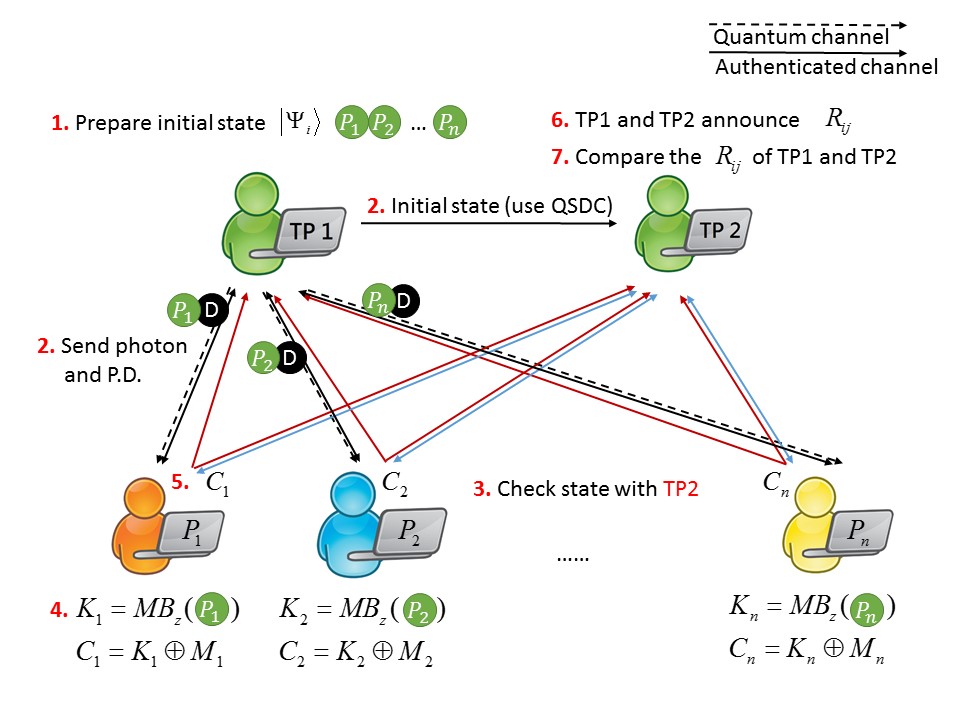}
\par\end{centering}
\caption{Proposed multiparty QPC protocol}
\end{figure}

The correctness of this protocol is based on the property of GHZ state
described in Section 3.1. Since we know the xoring value, $T_{ij}=K_{i}\oplus K_{j}$,
of the $i-$th and the $j-$th particles in a particular GHZ state,
we can calculate that $R_{ij}=T_{ij}\oplus C_{i}\oplus C_{j}$ $=T_{ij}\oplus K_{i}\oplus K_{j}\oplus M_{i}\oplus M_{j}=M_{i}\oplus M_{j}$.
Hence, if all bits in $R_{ij}$ are \textquoteleft 0\textquoteright ,
it means all bits in $M_{i}$ and $M_{j}$ are equal and the secret
information of $P_{i}$ and $P_{j}$ is identical.

\subsection{Stranger environment}

This protocol also can work on a stranger environment because the
involved users are communicating only on classical channels. In Step
3, in order to prevent from $TP_{1}$'s attacks, all participants
are communicating on classical channels, even though the communication
between each user and $TP_{2}$ will eventually detect it. For example,
in Step 3, $P_{1}$ randomly chooses the particles for checking and
announces the positions of those particles to the other participants
via classical channels. Now, if the information in the classical channel
is modified by an outsider, then the other participants will receive
wrong positions. In that case, they all measure the wrong photons
except $P_{1}$. Since the measurement results may not correspond
with the initial state with a high probability, $TP_{2}$ will detect
this fraud. For instance, suppose that the initial state of a particular
chosen position, $i$ $\left(1\leq i\leq m\right)$, is $\left|\varPsi_{1}\right\rangle =\frac{1}{\sqrt{2}}\left(\left|000\right\rangle +\left|111\right\rangle \right)$
$=\frac{1}{2}\left(\left|+++\right\rangle +\left|+--\right\rangle +\left|-+-\right\rangle +\left|--+\right\rangle \right)$.
If the correct photons in the state are measured, then the measurement
result will be $\left|000\right\rangle $ or $\left|111\right\rangle $
in Z-basis and $\left|+++\right\rangle $, $\left|+--\right\rangle $,
$\left|-+-\right\rangle $, or $\left|--+\right\rangle $ in X-basis.
However, if the checking position has been modified by an outsider
to the other position $j$ $\left(1\leq j\leq m\right)$ with the
initial state $\left|\varPsi_{3}\right\rangle =\frac{1}{\sqrt{2}}\left(\left|100\right\rangle +\left|011\right\rangle \right)$
$=\frac{1}{2}\left(\left|+++\right\rangle +\left|+--\right\rangle +\left|-+-\right\rangle +\left|--+\right\rangle \right)$,
then measures the photon in $i$, whereas the others will measure
the photons in $j$. Consequently the measurement result obtained
by $TP_{2}$ will become $\left|100\right\rangle $, $\left|011\right\rangle $,
$\left|000\right\rangle $, or $\left|111\right\rangle $ in Z-basis
and $\left|+++\right\rangle $, $\left|+--\right\rangle $, $\left|-+-\right\rangle $,
$\left|--+\right\rangle $, $\left|-++\right\rangle $, $\left|---\right\rangle $,
$\left|++-\right\rangle $, or $\left|+-+\right\rangle $ in X-basis
which will correspond to $\left|\varPsi_{1}\right\rangle $ with a
probability of 50\%. Hence, for $l$ initial states, the detection
rate is $1-\left(1/2\right)^{l}$ which is close to 1 if $l$ is large
enough. 

Since there are only classical channels between participants, there
may be DOS attack in Step 3 if the classical channels are frequently
disturbed. However, with a little modification, this sort of DOS attack
can be prevented. The modification is as follows. Instead of announcing
the positions of the chosen particles via classical channels, $P_{1}$
informs $TP_{2}$ the positions of the chosen particles via the authentication
channel, and $TP_{2}$ informs all the other participants that information
also via authentication channels. Similarly $P_{2}$ announces the
information via authentication channels in Step 3, too. With this
modification, no classical channel is used and hence the DOS attacker
cannot be successful. 

\subsection{Who is telling a lie}

As mentioned earlier that an individually dishonest TP could announce
a fake comparison result. However, since the other TP also does the
same comparison and announces the comparison result, participants
will eventually detect the inconsistency if one of the TPs is not
honest. Unfortunately, the current protocol cannot identify which
TP announced the fake comparison result. To identify the dishonest
TP, an arbitrated quantum signature protocol, e.g., \cite{17} can
be introduced to the proposed scheme with the help of a trusted arbitrator
as follows. In Step 2, instead of sending the information of the initial
GHZ states to $TP_{2}$, $TP_{1}$ signs the information of the initial
states via an arbitrated quantum signature for $TP_{2}$ and protects
the privacy of the content by using the keys between TPs and arbitrator.
Later, this information can be used by the arbitrator to identify
the TP who was telling a lie, because the arbitrator can use the signed
initial states to do the comparison and hence can identify the dishonest
TP.

\section{Security Analysis}

In this section, we show that our proposed protocol has several imperative
security properties, which are important for a secure QPC protocol.
This section contains two parts, the outsider attack (Section 4.1),
the insider attack (Section 4.2).

\subsection{Outsider attack}

After $TP_{1}$ sends all photons to each participant, all participants
and $TP_{1}$ perform public discussion to check outsider attack.
First, $TP_{1}$ announces the positions and bases of all decoy photons.
Later, each participant gets the measurement results by measuring
the corresponding decoy photons. Then, every participant sends back
the measurement results to $TP_{1}$. $TP_{1}$ checks the existence
of eavesdroppers by checking whether the measurement results are correct
or not.

Since the eavesdropper, Eve, does not know the positions and measurement
bases of the decoy photons, some well-known attacks such as intercept-resend
attack \cite{18}, correlation-elicitation attack \cite{19}, and
entanglement-measure attack \cite{20} can be detected via the checking
mechanism \cite{3}. For example, if Eve measures the decoy photon
$\left|0\right\rangle $ or $\left|1\right\rangle $ with Z-basis
$\left\{ \left|0\right\rangle ,\left|1\right\rangle \right\} $, she
will pass the public discussion. However, if Eve measures the decoy
photon $\left|0\right\rangle $ or $\left|1\right\rangle $ with X-basis
$\left\{ \left|+\right\rangle ,\left|-\right\rangle \right\} $, because
of the quantum property, the probability that she will be detected
is 50\%. Obviously, the probability that Eve chooses the wrong measurement
basis is 50\%. Therefore, the detection rate for each decoy photons
is 25\% (50\%\texttimes 50\%). For $l$ decoy photons (where $l$
is large enough), the detection rate is $1-\left(3/4\right)^{l}$
which is close to 1 if $l$ is large enough. Furthermore, since quantum
bits are transmitted only once in the proposed protocol, the Trojan
horse attack can be automatically prevented. Therefore, the proposed
protocol is free from any outsider attack.

\subsection{Insider attack}

In this sub-section, three cases of insider attack will be considered.
The first case discusses about the participants\textquoteright{} attack.
The second and third cases discuss the attack form $TP_{1}$ and $TP_{2}$,
respectively.

\subsubsection*{Case 1. Participants\textquoteright{} attack}

Suppose that Alice attempts to reveal Bob\textquoteright s secret.
$TP_{1}$ and $TP_{2}$ are individually dishonest TPs who will not
conspire with each other and with the participants. In this case,
if Alice tries to intercept the transmitted photon from $TP_{1}$
to Bob, she will be caught as an eavesdropper as discussed in Section
4.1. Therefore, the only possible way for Alice to obtain Bob\textquoteright s
private information is using her photon to extract Bob\textquoteright s
measurement result. If Alice knows the initial state, she could calculate
Bob\textquoteright s measurement result by the measurement result
of Alice\textquoteright s photon and the initial state. For example,
suppose the initial state is $\left|\varPsi_{1}\right\rangle =\frac{1}{\sqrt{2}}\left(\left|000\right\rangle +\left|111\right\rangle \right)$
$=\frac{1}{2}\left(\left|+++\right\rangle +\left|+--\right\rangle +\left|-+-\right\rangle +\left|--+\right\rangle \right)$.
If the measurement result of Alice $K_{A}$ is $0$, then she will
know the measurement result of Bob's $K_{B}$ is also 0. By knowing
$K_{B}$ and $C_{B}$, Alice can calculate the secret information
of Bob's $M_{B}$. However, since Alice does not know anything about
the initial state, it is impossible for her to perform this attack. 

\subsubsection*{Case 2. $TP_{1}$'s attack }

In the proposed protocol, the responsibility of $TP_{1}$ is to generate
initial states, inform $TP_{2}$ the initial states and compare the
private information. $TP_{1}$ may try to steal participants\textquoteright{}
secrets by using fake initial states instead of the official initial
states. However, in Step 3, $TP_{2}$ and all the participants work
together to check the correctness of the initial states, so if $TP_{1}$
use a fake initial state, he will be caught. For example, suppose
there are three participants. $TP_{1}$ generates $\left|000\right\rangle $
as the initial state and sends those particles to three participants
respectively as in Step 2, but $TP_{1}$ tells $TP_{2}$ a lie about
the initial state as $\left|\varPsi_{1}\right\rangle =\frac{1}{\sqrt{2}}\left(\left|000\right\rangle +\left|111\right\rangle \right)$
$=\frac{1}{2}\left(\left|+++\right\rangle +\left|+--\right\rangle +\left|-+-\right\rangle +\left|--+\right\rangle \right)$.
All participants use Z-basis to measure the photon and get the key
bit $K_{i}$. Since $TP_{1}$ knows that the photons all participants
received are $\left|0\right\rangle $, he/she will know the key bits
$K_{1}$ for each participants is \textquoteleft 0\textquoteright .
Then, each participant $P_{i}$ sends $C_{i}$ to $TP_{1}$. As the
result, $TP_{1}$ can easily calculate participants\textquoteright{}
secrets $M_{i}$. However, this attack of $TP_{1}$ will be detected
in Step 3, because if the participant $P_{2}$ chooses X-basis to
check, then the measurement results will not always be in $\left\{ \left|+++\right\rangle ,\left|+--\right\rangle ,\left|-+-\right\rangle ,\left|--+\right\rangle \right\} $
and hence the fake initial state of $TP_{1}$ will be detected.

\subsubsection*{Case 2. $TP_{2}$'s attack }

In the proposed protocol, the responsibility of $TP_{2}$ is to check
the correctness of the initial states and compare the private information
of each pair of users. $TP_{2}$ may try to steal participants\textquoteright{}
secrets by intercepting the transmitted photons from $TP_{1}$ to
participants. However, $TP_{2}$ will be caught in the eavesdropper
detection discussed in Section 4.1.

\section{Conclusion}

A new security problem about the trustworthiness of a TP, who could
announce a fake comparison result, in the state-of-the-art QPC protocols
is identified, which may cause the participants to believe in a wrong
comparison result. To explore further the problem, a new TP named
individually dishonest TP, is defined. Subsequently, a multiparty
QPC protocol, which provides a solution to detect the fake comparison
(or intermediate) result announced by a TP has been proposed. We argue
that the proposed protocol can also work in a stranger environment,
where there is no authentication channel or no pre-shared key between
each pair of participants. Moreover, the proposed protocol has been
shown to be secure against both the outsider and the insider attacks.

\section*{Acknowledgment}

We would like to thank the Ministry of Science and Technology of Republic
of China for financial support of this research under Contract No.
MOST 104-2221-E-006-102 -.

\section*{References}

\bibliographystyle{IEEEtran}
\bibliography{references}

\end{document}